\documentclass[aps,8pt,prd,twocolumn,preprintnumbers,amsmath,amssymb,nofootinbib,superscriptaddress,a4paper,showpacs]{revtex4-1}
\usepackage[colorlinks=true, pdfstartview=FitV, linkcolor=blue, citecolor=red, urlcolor=magenta]{hyperref}
\usepackage{graphicx}
\usepackage{latexsym}
\usepackage{amsmath}
\usepackage{amsfonts}
\usepackage{amssymb}
\usepackage{verbatim}
\usepackage{dcolumn}
\usepackage{amssymb}
\usepackage{bm}
\usepackage{float}
\usepackage{subfigure}
\usepackage[english]{babel}
\newcommand{\be}{\begin{equation}}
\newcommand{\ee}{\end{equation}}
\newcommand{\bea}{\begin{eqnarray}}
\newcommand{\eea}{\end{eqnarray}}

\begin{document}

\newcommand{\JPess}{
\affiliation{Departamento de F\'isica, Universidade Federal da Para\'iba, \\Caixa Postal 5008, 58059-900, Jo\~ao Pessoa, PB, Brazil}
}

\newcommand{\Quixada}{\affiliation{
Universidade Estadual do Cear\'{a}, Faculdade de Educa\c{c}\~{a}o, \\Ci\^{e}ncias e Letras do Sert\~{a}o Central, 63900-000, Quixad\'{a}, CE, Brazil
}}

\title{Effects of quantum corrections on the criticality and efficiency of black holes surrounded by a perfect fluid}

\author{V. B. Bezerra}
\email{valdir@fisica.ufpb.br}
\JPess
\author{I. P. Lobo}
\email{iarley\_lobo@fisica.ufpb.br}
\JPess
\author{J. P. Morais Gra\c ca}
\email{jpmorais@gmail.com}
\JPess
\author{Luis C. N. Santos}
\email{luis.santos@ufsc.br}
\JPess

%

\begin{abstract}
We study some properties of the extended phase space of a quantum-corrected Schwarzschild black hole surrounded by a perfect fluid. In particular we demonstrate that, due to the quantum correction, there exist first and second order phase transitions for a certain range of the state parameter of the perfect fluid, and we explicitly analyze some cases. Besides that, we describe the efficiency of this system as a heat engine and the effect of quantum corrections for different surrounding fluids.

\end{abstract}


\maketitle


\section{Introduction}

In the early seventies, Bekenstein proposed the idea that black holes have entropy and that this entropy is proportional to its area \cite{Bekenstein:1973ur,Bekenstein:1974ax}. At that time this idea was received with incredulity, since classical black holes cannot emit radiation, which means that they cannot have a well-defined temperature. It was Hawking who showed that, when quantum effects are taken into account, a black hole can indeed emit radiation, and this radiation can be understood as the radiation of the black hole as a perfect black body \cite{Hawking:1974sw}. Since then, several authors have studied such implications, and one of the most interesting is the possibility that black holes undergo phase transitions in the same way some thermodynamic systems do.

In daily life, the most usual transition we are aware is the phase transition of water becoming ice or vapor. Such transition occurs for some fixed temperature (isothermal curves) where the volume of the system varies for a fixed pressure. It can be modeled, as a first approximation, by the Van der Wall equation of state. One can be led to ask if some analogue procedure can also happen for a black hole, and for such task it is compulsory to properly define what is the pressure of the black hole. This issue is not straightforward, since for the most simple black hole, based on the Schwarzschild metric, a proper definition of pressure appears to be lacking. This issue was solved by Kastor, Ray and Traschen \cite{Kastor:2009wy} by the use of the cosmological constant as a pressure term and the identification of the black hole's mass with the enthalpy. Recently, Kubiznak and Mann \cite{Kubiznak:2012wp} were able to study the $P-V$ diagram of such thermodynamic black hole. Following these lines, one is able to identity the Schwarzschild anti-de Sitter spacetime as a full thermodynamic system.

As understood by Hawking, to properly study the thermodynamics of black holes one needs to take into account quantum effects, and to be consistent with this premise one should also take into account the back-reaction of these quantum effects on the metric. This procedure can be performed in a different number of ways, and a particular deformation of the Schwarzschild metric has been studied by Kazakov and Solodukhin \cite{Kazakov:1993ha}, and recently extended for the Schwarzschild metric surrounded by quintessence by Shahjalal \cite{Shahjalal:2019pqb}, extending the solution found by Kiselev \cite{Kiselev:2002dx}.

In fact, Kiselev's solution can be applied to the case of any perfect fluid surrounding the black hole. Using this technique, we shall recover the quantum corrected version of this metric found in \cite{Shahjalal:2019pqb}, and study the extended phase space of this spacetime as a thermodynamic system, in which we follow the procedure of \cite{Shahjalal:2019pqb} and treat the perfect fluid as generator of the pressure term. This choice is consistent when the field is identified with a negative cosmological constant, but one should take care when one is dealing with arbitrary fields, such as quintessence, phantom matter and others.

As we shall see, the introduction of the quantum correction is important since, usually, to obtain phase transitions on a non-rotating black hole in General Relativity one needs to introduce charges in the black hole, so that the spacetime is described by the Reissner-Nordstr\"{o}m anti-de Sitter metric instead of the Schwarzschild anti-de Sitter metric. If, instead, one introduces quantum corrections, then phase transitions can occur even in the absence of charges. Such phase transitions have been explored in several contexts and gravitational theories, for instance assuming non-linear electrodynamics \cite{Simovic:2019zgb,Dayyani:2017fuz}, a cloud of strings environment \cite{Toledo:2019szg,MoraisGraca:2018ofn,Ma:2019pya}, the effect of quark-gluon matter \cite{Critelli:2017oub}, alternative gravitational theories \cite{Brihaye:2018nta,Ovgun:2017bgx, Ma:2017pap,Upadhyay:2017fiw,Fernando:2016sps,Hendi:2015kza,Dehghani:2014caa,Mo:2014mba}, quantum gravity phenomenology \cite{Hendi:2016njy}, accelerating black holes \cite{Liu:2016uyd} and other alternative formulations \cite{Majhi:2016txt}. Besides these, some properties of the thermodynamics of a quantum-corrected Schwarzschild black hole has been studied in \cite{Kim:2012cma}.

Also, the presence of a pressure term allows us to consider thermodynamic cycles and the behavior of such black hole as a heat engine. We shall analyze the effect of quantum corrections on the efficiency of some processes for different perfect fluids surrounding the black hole. This framework was first idealized by Johnson \cite{Johnson:2014yja} and has been explored in different contexts \cite{Balart:2019uok,Zhang:2018hms,Fernando:2018fpq,EslamPanah:2018ums,Mo:2018hav,Rosso:2018acz,Zhang:2018vqs,Chakraborty:2017weq,Wei:2017vqs,Hendi:2017bys,Xu:2017ahm,Liu:2017baz,Hennigar:2017apu,Chakraborty:2016ssb,Zhang:2016wek,Bhamidipati:2016gel,Johnson:2015fva,Johnson:2015ekr,Setare:2015yra}.

This paper is organized in the following way. In section II we will briefly review the metric for the quantum-corrected black hole surrounded by a perfect fluid. In section III we will study the criticality of such objects, for different choices of matter fields. We will look for phase transitions of the first and second kinds. In section IV we will consider these black hole as heat engines, and describe their efficiency for different fluids. Finally, in section V, we will make some comments about the results that we found in this work.


\section{Quantum corrected black hole surrounded by quintessence}

\subsection{Quantum corrected Schwarzschild metric}

The classical dynamics of a field in a gravitational background is given by the Einstein-Hilbert action plus the matter action, namely

\begin{equation}
    S = \int d^4x \sqrt{^4|g|} \left( \frac{1}{16 \pi G_N} R^{(4)} + \mathcal{L}_{int}\right),
\end{equation}

\noindent
where $R^{(4)}$ is the Ricci scalar in four dimensions. Usually, one does not consider the back-reaction of the perturbations of the fields in the metric, so that the metric can be calculated by the Einstein equations with an energy-momentum tensor obtained from the matter fields. The consideration of the back-reaction of perturbations of the fields (both the matter and the metric itself) in the metric is a highly non-trivial and non-linear problem. A method to solve such issue for scalar fields has been developed by Kazakov and Solodukhin \cite{Kazakov:1993ha}, taking into account that one can reduce the problem to the effective two-dimensional dilaton gravity. To follow this procedure, we start writing an arbitrary spherically symmetric metric in the form

\begin{equation}
ds^2 = g_{\alpha\beta}(z) dz^\alpha dz^\beta - r^2(z) (d\theta^2 + \text{sin}^2 \theta d\phi^2),    
\end{equation}

\noindent
where the coordinates $(z^0, z^1, \theta, \phi)$ covers the whole spacetime. One can choose the two-dimensional metric $g_{\alpha\beta}$ as conformally flat, such that the metric takes the form

\begin{equation}
ds^2 = e^\sigma(z^+, z^-) dz^+ dz^- - r^2(r^+, r^-) (d\theta^2 + \text{sin}^2 \phi),
\end{equation}

\noindent
where the coordinates $(z^+, z^-)$ are null. After integrating out the angular coordinates, the four dimensional Einstein-Hilbert action reduces to

\begin{equation}
S_{EH} = - \frac{1}{8 G_N} \int d^2z \sqrt{^2|g|} [r^2 R^{(2)} - 2(\nabla r)^2 + 2],
\label{action1}
\end{equation}

\noindent
where the determinant and the Ricci scalar are now calculated from the two-dimensional metric $g_{\alpha \beta}$. Introducing the dilaton field $\phi = \ln(r^2/G_N)$, equation (\ref{action1}) takes the form of the dilaton gravity,

\begin{equation}
S_{EH} = - \frac{1}{8 G_N} \int d^2z \sqrt{|g|} \left[e^\phi\left(R - \frac{1}{2}(\nabla \phi)^2 + U(\phi)\right)\right],   
\label{action2}
\end{equation}

\noindent
where $U(\phi) = 2/G_N$. One now can use all the methods of two-dimensional dilaton gravity to solve the metric for action ($\ref{action2}$). Also following the approach adopted in \cite{Kazakov:1993ha}, let us consider the generalized action with an arbitrary dilaton potential $U(r)$,

\begin{equation}
    S_{EH} = -\frac{1}{8} \int d^2z \sqrt{|g|} \left[r^2 R^{(2)} - 2(\nabla r)^2 + \frac{2}{G_N}U(r)\right],
\end{equation}

\noindent
where $r \rightarrow r/\sqrt{G_N}$. A solution for this action is a Schwarzschild-like metric,

\begin{equation}
    ds^2 = - A(r) dt^2 + A(r)^{-1} dr^2 + r^2 d \Omega^2,
    \label{smetric}
\end{equation}

\noindent
where 

\begin{equation}
    A(r) = - \frac{2M}{r} + \frac{1}{2} \int^r U(\rho) d \rho,
\end{equation}

\noindent
and the non-trivial issue is to find a potential $U(\rho)$ such that the theory is renormalizable. The details for such procedure can be found in \cite{Kazakov:1993ha}, and the renormalized potential is given by

\begin{equation}
U(r) = \frac{r}{\sqrt{r^2 - 16 G_R}},    
\end{equation}

\noindent
where $G_R$ is the renormalized Newton constant, and $r > 4 G_R$. Defining a new parameter $a = 4 \sqrt{G_R} \leq r$, we obtain the result

\begin{equation}
A(r)=-\frac{2M}{r}+\frac{\sqrt{r^2-a^2}}{r}.
\end{equation}


\subsection{Quantum corrected black hole surrounded by a perfect fluid}

Following Kiselev \cite{Kiselev:2002dx}, one can define a spherically symmetric perfect fluid as a background with the energy-momentum tensor,

\begin{eqnarray}
T^t_t = T^r_r = \rho, \\
T^\theta_\theta = T^\phi_\phi = - \frac{1}{2}\rho (3 \omega - 1),
\end{eqnarray}

\noindent
where 

\begin{equation}\label{energy-density}
\rho = -\frac{c}{2} \frac{3\omega}{r^{3(1+\omega)}},
\end{equation}

\noindent
with $c$ being an arbitrary constant, and $\omega$ being the parameter of the equation of state $p = \omega \rho$. For exotic fluids with $\omega<0$, having $c > 0$ guarantees the positiveness of the energy density. With this energy-momentum tensor, one can look for the solution of a quantum-corrected black hole surrounded by such perfect fluid. The obtained result is the Schwarzschild-like metric, given by eq. (\ref{smetric}), with

\begin{equation}
A(r)=-\frac{2M}{r}+\frac{\sqrt{r^2-a^2}}{r}-\frac{c}{r^{3\omega +1}}.
\end{equation}

\noindent
Our aim in the following section is to work out the critical conditions for this space-time. 


\section{Criticality of quantum corrected black hole}
Considering the quantum corrected spherically symmetric black hole surrounded by a perfect fluid, we now move the description of the thermodynamic system in the so called extended phase space, in which the black hole's mass is the enthalpy, the area of the event horizon is the entropy and the constant, $c$, of the perfect fluid plays the role of a pressure term. This approach has also been named as black hole chemistry \cite{Kubiznak:2016qmn}, and allows us to study phenomena like phase transitions and criticality, besides to treat the black hole as a heat engine with a certain efficiency.
\par
We start by searching for the relation between the black hole's mass and the location of its horizon. In order to do so, we need to solve the equation:
\begin{equation}
A(r_+)=0,
\end{equation}
which implies
\begin{equation}
M=\frac{1}{2}\sqrt{r_+^2-a^2}-\frac{c}{2\, r_+^{3\omega}}.
\end{equation}

In \cite{Shahjalal:2019pqb}, Shahjalal has shown that the entropy of a quantum-corrected Schwarzschild black hole surrounded by quintessence is given by the Bekenstein entropy, so one can consider  
\begin{equation}\label{entropy}
S=\pi r_+^2.
\end{equation}

The term related with the pressure is a little tricky. Usually, one chooses the pressure as proportional to the cosmological constant, since this term appears in the first law as conjugate to the volume, like $V dP$. But now the terms on the first law will depend on the kind of perfect fluid field one chooses, i.e., will depend on the coefficient $\omega$. \cite{Shahjalal:2019pqb} has defined the pressure such that, when $\omega = -1$, this quantity is compatible with the usual procedure in Schwarzschild anti-de Sitter space, and in this  work we will follow his idea. So, we will choose
\begin{equation}
P=-\frac{3}{8\pi} c.
\end{equation}
The black hole mass plays the role of the enthalpy
\begin{equation}\label{mass-entropy}
M(S,P)=\frac{4\pi}{3}\left(S/\pi\right)^{-3\omega/2}P+\frac{1}{2}\sqrt{S/\pi-a^2}.
\end{equation}

As long as we already defined the pressure, the first law of thermodynamics is consistent if the volume of the system is given by

\begin{equation}\label{therm-vol}
V=\frac{4}{3}\pi r_+^{-3\omega}.
\end{equation}

Note that, for $\omega = -1$, this is the usual volume in plane geometry, also called ``areal volume''. For quintessence fluids, the volume will increases while increasing the area. However, in principle, the formulation due to Kiselev can be applied to any perfect fluid, and so it allows us to work with volumes such that they decrease as we increase the area. To illustrate, we will also work out the case of a radiation field. For some examples of other thermodynamic volumes, see for instance \cite{Altamirano:2014tva,Hennigar:2016gkm}.

The temperature of the system can be calculated as

\begin{equation}\label{temperature}
T=\frac{\partial M(S,P)}{\partial S}\bigg\rvert_{P}=-2\omega P\left(S/\pi\right)^{-(3\omega+2)/2}+\frac{1}{4\pi\sqrt{S/\pi-a^2}}.
\end{equation}

\noindent
This equation can be inverted so that we obtain an equation for the pressure as function of the radius and temperature, as follows
\begin{equation}
P(r_+,T)=\frac{1}{8\pi}\frac{r_+^{3\omega+2}\left(1-4\pi T\sqrt{r_+^2-a^2}\right)}{\omega\sqrt{r_+^2-a^2}}.
\label{Pressure}
\end{equation}

This allow us to look for a critical temperature. The conditions for the existence of critically are given by

\begin{eqnarray}
\begin{cases}
\partial P(r_c,T_c)/\partial r_+\big\rvert_{T}=0,\\
\partial^2 P(r_c,T_c)/\partial r_+^2\big\rvert_{T}=0,
\end{cases}
\end{eqnarray}

\noindent
and the critical temperature and critical radius are given by, respectively, 
\begin{equation}
T_c=\frac{\sqrt{3}}{18\pi}\frac{(3\omega+1)}{(3\omega+2)}\frac{\sqrt{3\omega+1}}{a},
\end{equation}

\noindent
and

\begin{equation}
r_c=\frac{\sqrt{(3\omega+1)(3\omega+4)}}{3\omega+1}a.
\end{equation}

The above results can be inserted in equation (\ref{Pressure}), and we finally obtain a critical pressure, 
\begin{equation}\label{crit-press}
P_c=\frac{\sqrt{3}}{72\pi}\frac{a^{3\omega+1}(3\omega+1)^{-(3\omega+1)/2}(3\omega+4)^{(3\omega+4)/2}}{\omega(3\omega+2)}.
\end{equation}
Therefore, we only have criticality if $\omega>-1/3$ and $\omega\not= 0$, and additionally $a\not=0$. Let us note that, usually, one adds a Reissner-Nordstr\"{o}m term in the metric to study the criticality of the system, since the Schwarzschild black hole by itself does not have a critical temperature in the $P-V$ plane. The introduction of quantum-correction and quintessence allow us to find criticality even without a Reissner-Nordstr\"{o}m term in the metric.  

Defining the quantity 
\begin{equation}
v_c\doteq\frac{3}{2}\frac{\omega(3\omega+1)}{3\omega+4} r_c^{-(3\omega+2)},
\end{equation}
with a little algebra, one can find a dimensionless golden relation for the quantities $P_c, T_c$ and $v_c$, given by
\begin{equation}
\frac{P_c\, v_c}{T_c}=\frac{3}{8},
\end{equation}
which is similar to relations found in the context of the Van der Waals equation and on the study of criticality of Reissner-Nordstrom AdS black hole \cite{Kubiznak:2012wp}. This relation depends on the kind of fluid that surrounds the black hole, but does not depend on the scale of the quantum correction.
\par
To properly study the relation between pressure and volume, we will define the  dimensionless variables $t=a\, T$, $x=r_+/a$ and $p=P/a^{3\omega+1}$. Then, equation ($\ref{Pressure}$) can be recast as
\begin{equation}\label{px}
p(x,t)=\frac{1}{8\pi\omega}x^{3\omega+2}\left(\frac{1}{\sqrt{x^2-1}}-4\pi t\right).
\end{equation}
Besides that, the Gibbs free energy is defined as
\begin{equation}\label{gibbs1}
G=M-TS=\frac{r_+^2-2a^2}{\sqrt{r_+^2-a^2}}+\frac{2\pi}{3}(3\omega+2)\frac{P}{r_+^{3\omega}},
\end{equation}
where $r_+=r_+(T,P)$ is a function of the temperature and the pressure.
\par
If we define $g=G/a$, Eq.(\ref{gibbs1}) can be recast as
\begin{equation}
g(x,p)=\frac{x^2-2}{4\sqrt{x^2-1}}+\frac{2\pi}{3}(3\omega+2)\frac{p}{x^{3\omega}},
\end{equation}
and the temperature as 
\begin{equation}\label{temperature1}
t(s,p)=-2\omega\, p\, \tilde{s}^{-(3\omega+2)/2}+\frac{1}{4\pi\sqrt{\tilde{s}-1}},
\end{equation}
where $\tilde{s}=S/\pi a^2$.
\par 
Considering these redefinitions, for some fluids surrounding the black hole, we depict $p-x$ and $g-t$ diagrams for fixed temperature and pressure, respectively.


\subsection{Radiation fluid}
\begin{figure}[H]
\centering
\includegraphics[width=7.5cm,height=5cm]{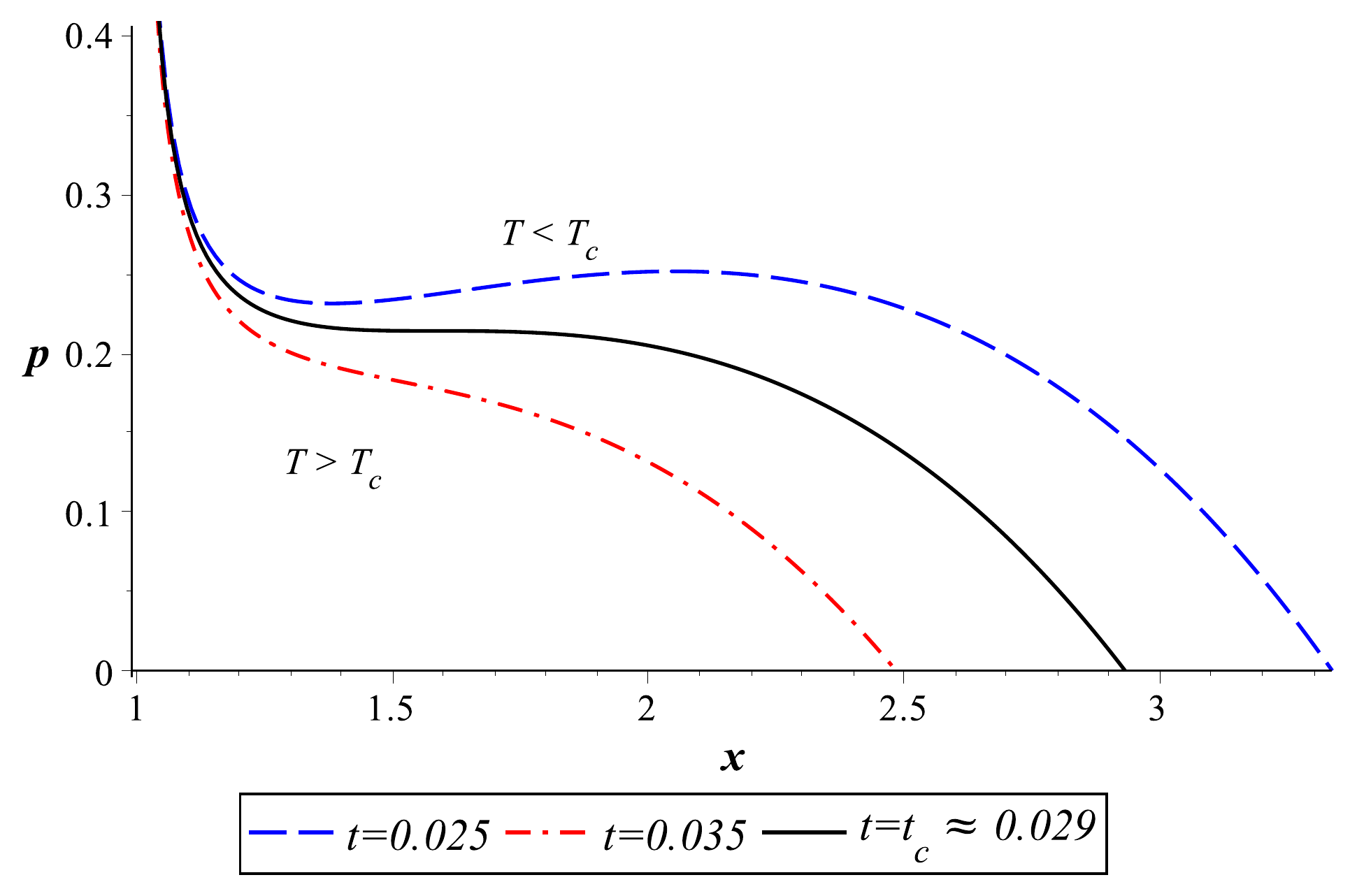}
\caption{$p-x$ diagram for the radiation case $\omega=1/3$. Here $V\propto r^{-1}$.}\label{crit:rad}
\end{figure}

As a first case, we depict the behavior of isotherms of the radiation case in Fig.(\ref{crit:rad}). Using $\omega=1/3$ in Eq.(\ref{px}) we see that the pressure diverges when $x\rightarrow 1$, i.e., when the horizon radius approaches the length scale of the quantum correction. As can be seen, the black hole becomes electrically neutral for a limited value of the horizon $x_n$,
\begin{equation}
x_n=\sqrt{1+\frac{1}{16\pi^2 t^2}}.
\end{equation}
This maximum radius for keeping the black hole electrically charged increases with the reduction of the temperature, as described in Fig.(\ref{crit:rad}).
\par
We also verify the presence of criticality on the isotherm represented by the black (solid) curve. In this case, the variable conjugate to $P$ is proportional to $r^{-1}$, which not even grows with the horizon radius. Thus, it does not represent a quantity that can be identified with a physical volume. And a natural consequence of this statement is that $P$ can no longer be interpreted as a physical pressure. In fact, in this radiation case, we must rely on the more primitive assumption regarding $P$ as describing simply the electric charge of the black hole and the volume as the electric potential. Since this behavior will occur for any fluid with a positive value of $\omega$, we shall consider from now on in this section only cases with $\omega<0$. 

\begin{figure}[H]
\centering
\includegraphics[scale=0.45]{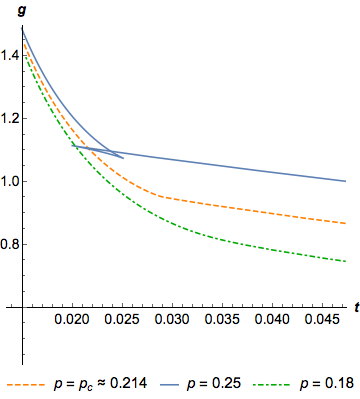}
\caption{$g-t$ diagram for the radiation case $\omega=1/3$.}\label{gibbs:rad}
\end{figure}

The critical behavior can also be seen in the $g-t$ diagram for isobaric curves, where the first order phase transition is evident from the discontinuity of the derivative $\partial g/\partial t|_p$ in the solid (blue) curve of Fig.(\ref{gibbs:rad}).

\subsection{Exotic fluids}

\begin{figure}[H]
\centering
\subfigure[ref1][\, $\omega=-1/6$]{\includegraphics[width=7cm,height=5cm]{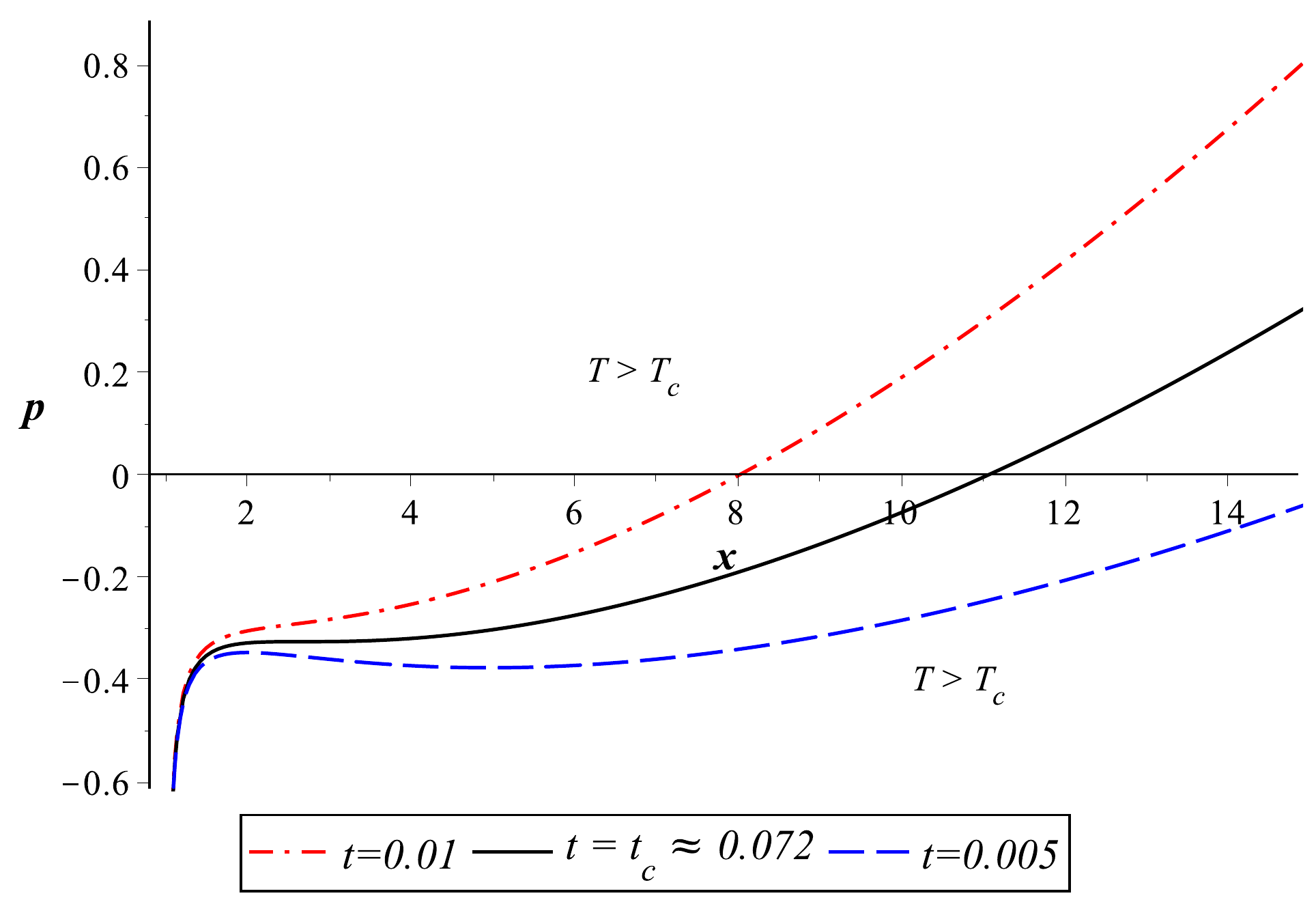}\label{crit:-1/6}}
\qquad
\subfigure[ref2][\, $\omega=-2/3$]{\includegraphics[width=7cm,height=5cm]{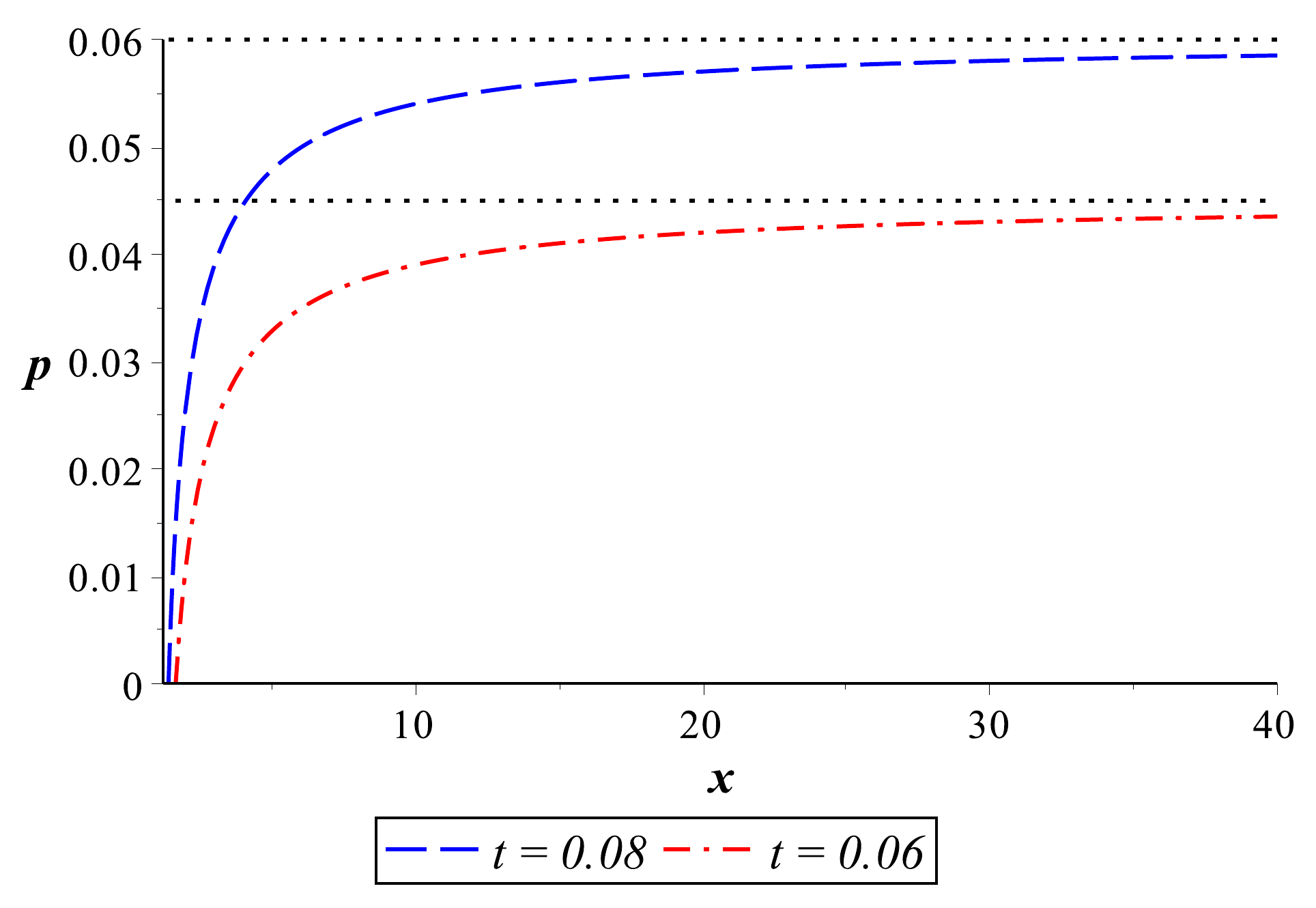}\label{crit:-2/3}}
\caption{$p-x$ diagram for the cases $\omega=-1/6$ and $\omega=-2/3$. In these cases, $V\propto \sqrt{r_+}$ and $V\propto r_+^2$, respectively.}
\end{figure}
\begin{figure}[H]
\centering
\subfigure[][\, $\omega=-1/6$]{\includegraphics[scale=0.4]{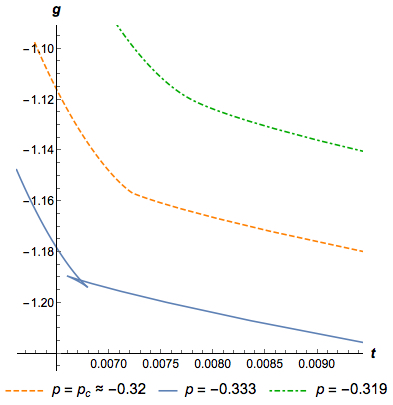}\label{gibbs:-1/6}}
\qquad
\subfigure[][\, $\omega=-2/3$]{\includegraphics[scale=0.4]{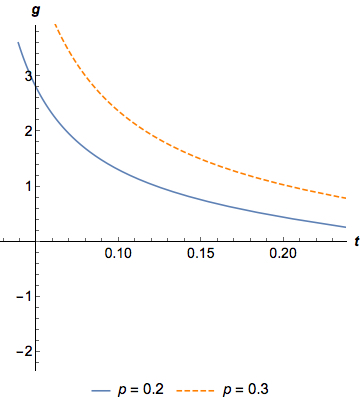}\label{gibbs:-2/3}}
\caption{$g-t$ diagram for the cases $\omega=-1/6$ and $\omega=-2/3$.}
\end{figure}

The late-time cosmic acceleration of the universe bounds the equation of state of fluid responsible for this acceleration to have $\omega<-1/3$ \cite{Sahni:2004ai}, but as we stated before, we only verify criticality for the opposite brach $\omega>-1/3$.  However, in order to illustrate the critical behavior of our black hole for exotic fluids, we analyze the latter case by assuming $\omega=-1/6$ in Figs.(\ref{crit:-1/6}) and (\ref{gibbs:-1/6}). This particular choice captures the overall nature of the $p-x$ and $g-t$ diagrams for any case in which $\omega\in (-1/3,0)$. For negative values of $\omega$, the critical pressure is always negative as can be verified from Eq.(\ref{crit-press}).
\par
For $\omega=-2/3$, the volume behaves like an area, $V\propto r^2$. In this case, we do not verify the critical behavior, as can be seen in Figs.(\ref{crit:-2/3}) and (\ref{gibbs:-2/3}), but we can see an upper bound on the pressure:
\begin{equation}
p_{\text{upper}}=\frac{3}{4}t.
\end{equation}

\begin{figure}[H]
\centering
\includegraphics[scale=0.4]{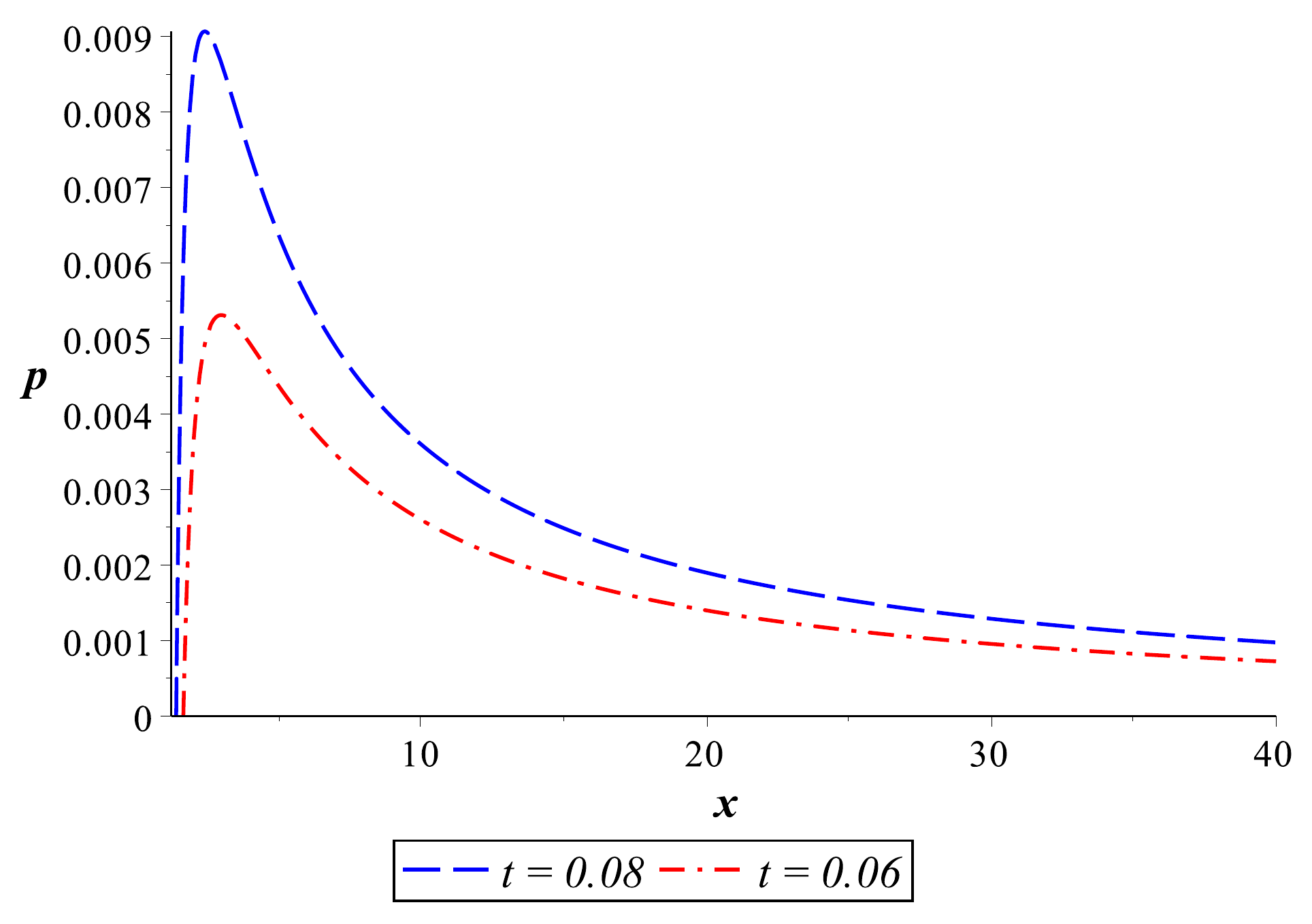}
\caption{$p-x$ diagram for the anti-de Sitter case, $\omega=-1$. Here $V\propto r^{3}$.}\label{crit:-1}
\end{figure}

The case $\omega =-1$ is most described in the literature, i.e., with the perfect fluid being a cosmological constant, furnishing a quantum corrected anti-de Sitter black hole. Here, the thermodynamic volume behaves as the usual geometric one with $V\propto r^3$. The pressure reaches a maximum value and goes to zero as the volume grows as can be seen in Fig.(\ref{crit:-1}).

\begin{figure}[H]
\centering
\includegraphics[scale=0.45]{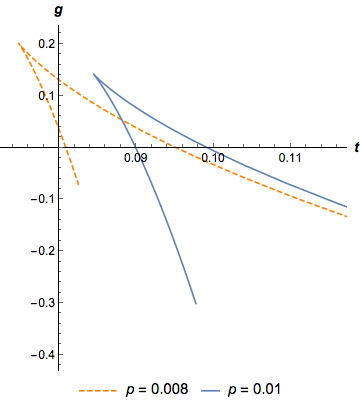}
\caption{$g-t$ diagram for the anti-de Sitter case, $\omega=-1$.}\label{gibbs:-1}
\end{figure}

From Fig.(\ref{gibbs:-1}), we see the presence of a minimal temperature $t_0$, which can be measured from Eq.(\ref{temperature1}). Solving $\partial t/\partial \tilde{s}|_{p}=0$, we find the entropy
\begin{equation}\label{s0}
\tilde{s}_0=B(p)
\end{equation}
where
\begin{eqnarray}\label{Bop}
B(p)=1+\frac{\left(3\sqrt{3}\sqrt{432\pi^2p^2-1}+108\pi p\right)^{1/3}}{24\pi p}\\
+\frac{1}{8\pi p \left(3\sqrt{3}\sqrt{432\pi^2p^2-1}+108\pi p\right)^{1/3}}.\nonumber
\end{eqnarray}
and substituting the result back into (\ref{temperature1}), we find
\begin{equation}
t_0(p)=2p \sqrt{B(p)}+\frac{1}{4\pi \sqrt{B(p)-1}},
\end{equation}  

\begin{figure}[H]
\centering
\includegraphics[scale=0.45]{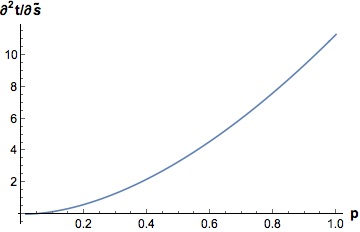}
\caption{Behavior of $\partial t/\partial \tilde{s}$, at $\tilde{s}=\tilde{s}_0$, as a function of the pressure $p$.}\label{2deriv}
\end{figure}

The second derivative at this point reads
\begin{equation}
\frac{\partial^2 t (s_0)}{\partial \tilde{s}^2}\bigg\rvert_{p}=-\frac{p}{2B(p)^{3/2}}+\frac{3}{16\pi\left(B(p)-1\right)^{5/2}},
\end{equation}
which is an ever positive function of $p$, as can be seen in Fig.(\ref{2deriv}), which characterizes a minimum point.
For example, for the values considered in Fig.(\ref{gibbs:-1}), we have $t_0(0.01)\approx 0.084$ and $t_0(0.008)\approx 0.074$.

This is similar to a Hawking-Page phase transition that separates two phases of the black hole \cite{Hawking:1982dh,Banerjee:2011au}. The concave curve represents smaller values of the horizon radius in comparison to the convex curve. And since for the anti-de Sitter case, the mass grows with the horizon radius, the first phase (concave portion) corresponds to a lower mass and the second one (convex portion) to a higher mass black hole, respectively.
\par
The threshold mass can be easily calculated from Eq.(\ref{mass-entropy}) as a function of the pressure as 
\begin{equation}
m_0(p)\doteq m|_{\tilde{s}=\tilde{s}_0}=\frac{4\pi}{3}pB(p)^{3/2}+\frac{1}{2}\sqrt{B(p)-1},
\end{equation}
where, as usual, we define the dimensionless quantity $m=M/a$. Such transition is of second order and occurs due to a discontinuity of the heat capacity that we shall analyze in the next subsection.

\subsection{Heat Capacity and second order phase transitions}
The heat capacity measures the amount of heat necessary to raise to temperature of the system by a given amount \cite{reichl}. The black hole heat capacity at constant pressure can be calculated from Eq.(\ref{temperature}) as 
\begin{eqnarray}
C_P=T\frac{\partial S}{\partial T}\bigg\rvert_{P}\nonumber\\
=\frac{16\pi^2\tilde{S}(\tilde{S}-a^2)(\omega P \sqrt{\tilde{S}-a^2}-\frac{1}{8\pi}\tilde{S}^{(3\omega+2)/2})}{8\pi P\omega (3\omega+2)(\tilde{S}-a^2)^{3/2}-\tilde{S}^{(3\omega+4)/2}},
\end{eqnarray}
where $\tilde{S}=S/\pi$. According to (\ref{entropy}) and (\ref{therm-vol}), if the volume is constant, then the entropy is also constant, which means that isochoric are also adiabatic processes.
In particular, this also implies that the heat capacity at constant volume is null
\begin{equation}
C_V=T\frac{\partial S}{\partial T}\bigg\rvert_{V}=0.
\end{equation}

\begin{figure}[H]
\centering
\includegraphics[width=7cm,height=5.5cm]{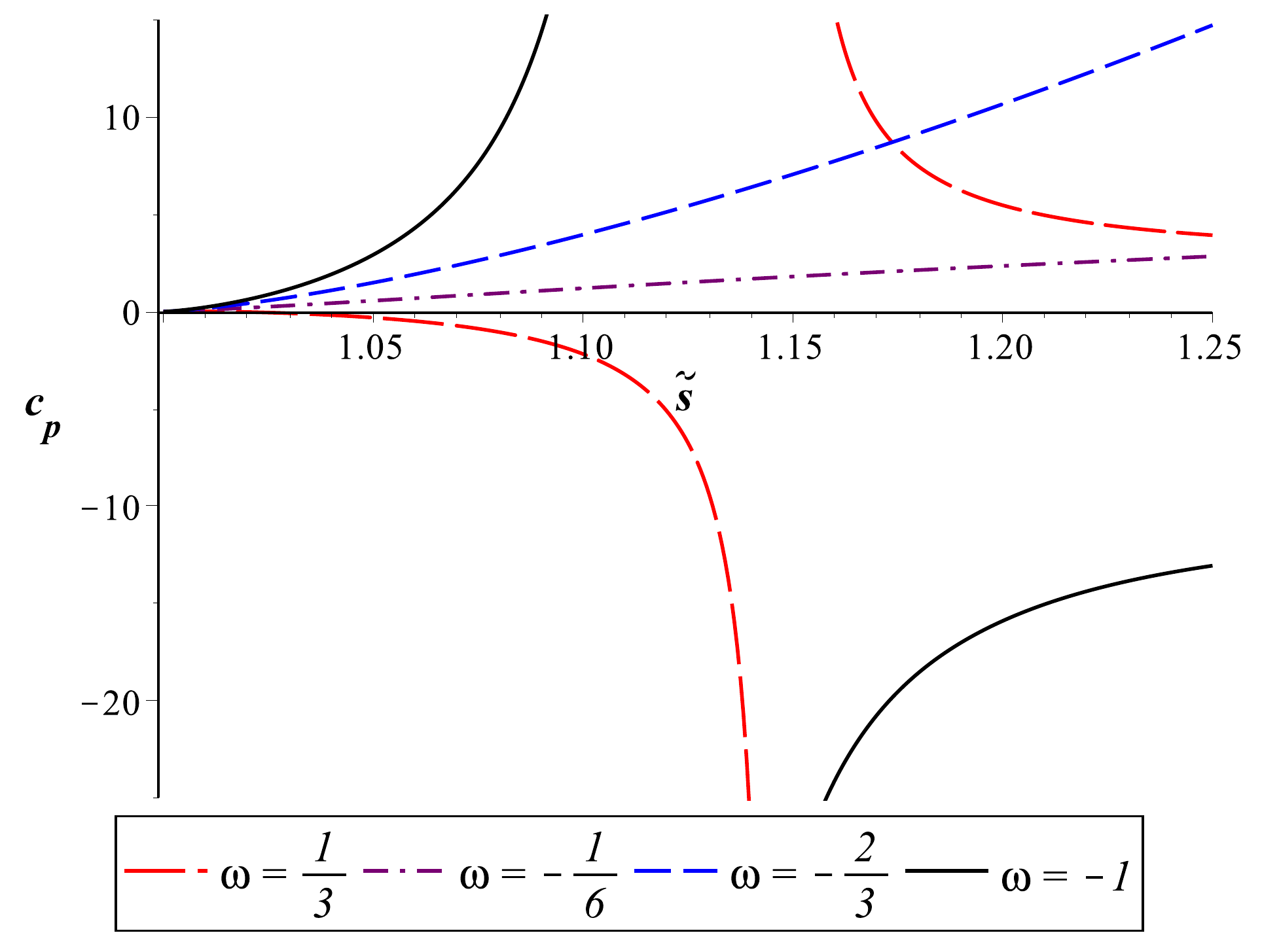}
\caption{$c_p-\tilde{s}$ diagram, assuming $p=1$, for $\omega=1/3$, $\omega=-1/6$, $\omega=-2/3$ and $\omega=-1$.}
\label{HC-2d}
\end{figure}

Redefining to adimensional variables $\tilde{s}=\tilde{S}/a^2$, $p=P/a^{(3\omega+1)/2}$ and $c_p=C_P/a^2$, we have the adimensional heat capacity at constant pressure
\begin{equation}\label{ad-heat}
c_p=\frac{16\pi^2\tilde{s}(\tilde{s}-1)(\omega \,p \, \sqrt{\tilde{s}-1}-\frac{1}{8\pi}\tilde{s}^{(3\omega+2)/2})}{8\pi \, p\, \omega (3\omega+2)(\tilde{s}-1)^{3/2}-\tilde{s}^{(3\omega+4)/2}},
\end{equation} 
and we can plot it for different surrounding fluids in Fig.(\ref{HC-2d}).

\begin{figure}[H]
\centering
\includegraphics[scale=0.4]{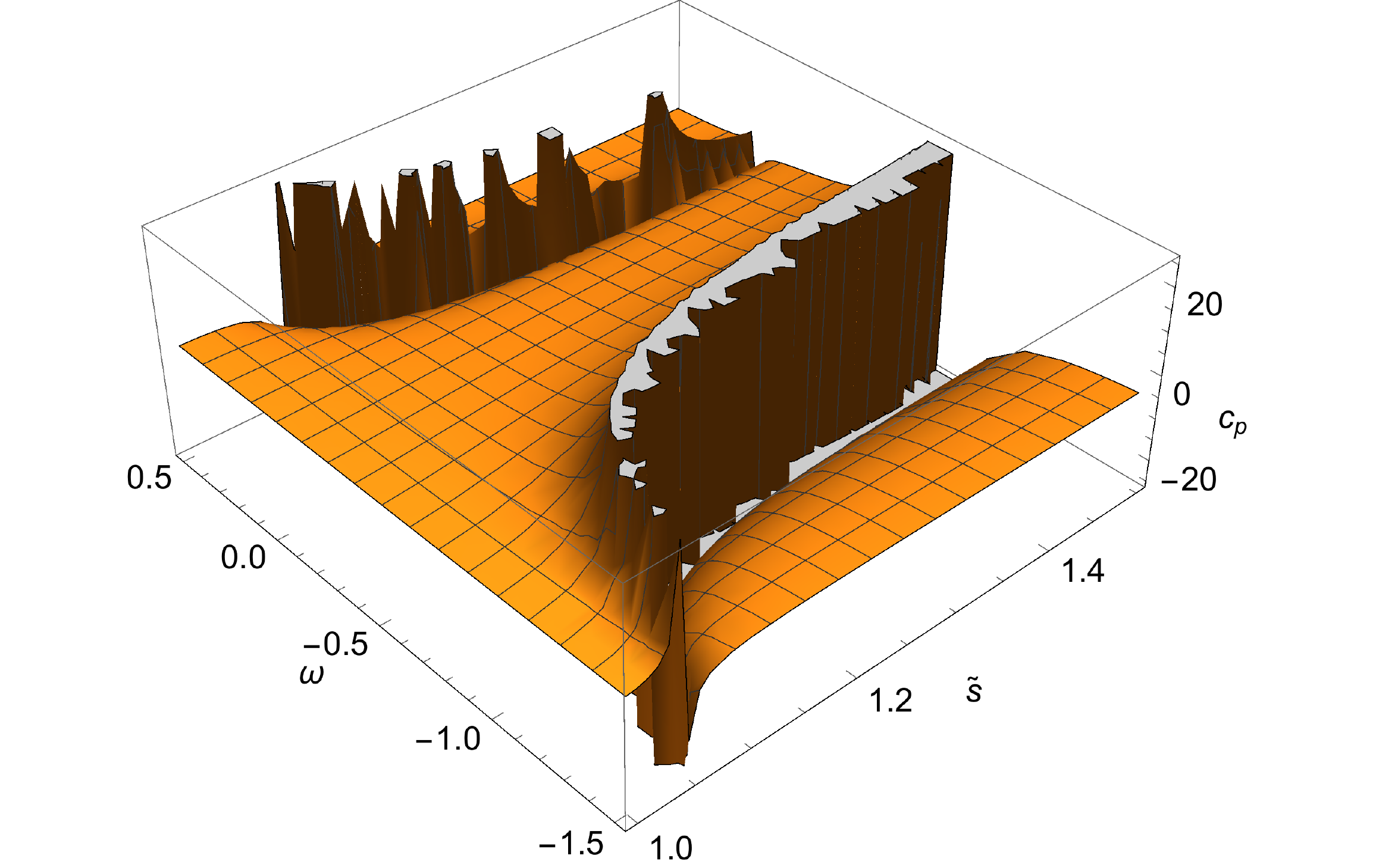}
\caption{$c_p-\tilde{s}-\omega$ diagram, assuming $p=1$.}
\label{HC-3d}
\end{figure}

As we can see, for $\omega=1/3$ and $\omega=-1$, we have second order phase transitions due to discontinuities in the heat capacity at constant pressure. However, they are absent for $\omega=-1/6$ and $\omega=-2/3$. Aiming to further analyze such differences for different values of $\omega$, we also depicted the three-dimensional behavior for a range of perfect fluid parameters $\omega\in[-3/2,1/2]$ in Fig.(\ref{HC-3d}) and we verify that for a certain range of the parameter $\omega$ there are no second order phase transitions. In fact, the denominator of Eq.(\ref{ad-heat}) never diverges if $\omega(3\omega+2)\leq0$, which means that for $\omega\in [-2/3,0]$ this system does not present second order phase transitions, for example the cases $\omega=-2/3$ and $\omega=-1/6$ in Fig.(\ref{HC-2d}).

Comparing to results of the last subsection, we see that for $\omega=-1$, the heat capacity diverges at $\tilde{s}=\tilde{s}_0$ given by Eqs.(\ref{s0}) and (\ref{Bop}).

\section{Black hole as heat engine}

Thermodynamics have started as a theoretical tool to study the efficiency of heat engines, and it is remarkable that the second law of thermodynamics, as developed by Carnot around 1820, has been first idealized in the framework of the old caloric theory of heat. But the thermodynamics of black holes took more than forty years, since its development in the seventies, to study black holes as thermal machines capable of producing work as the consequence of absorbing heat.  

Usually, a heat engine works between two reservoirs of temperature, a hot one and a cold one, and heat flows between these reservoirs. The product of the machine is work and the byproduct is the waste of some heat, i. e., not all heat is converted in work. The efficiency of the machine is defined by

\begin{equation}
    \eta = \frac{W}{Q_H}
    \label{efficiency}
\end{equation}

\noindent
where $W$ is the work produced and $Q_H$ is the amount of heat coming from the hot reservoir. The most efficient heat engine performs what is known as a Carnot cycle, constructed by two isothermals and two adiabatics. The most important fact about the Carnot cycle is that the efficiency of the heat engine depends only on the temperature of the reservoirs, and it is given by

\begin{equation}
    \eta = 1 - \frac{T_C}{T_H},
\end{equation}

\begin{figure}[H]
\centering
\includegraphics[scale=0.35]{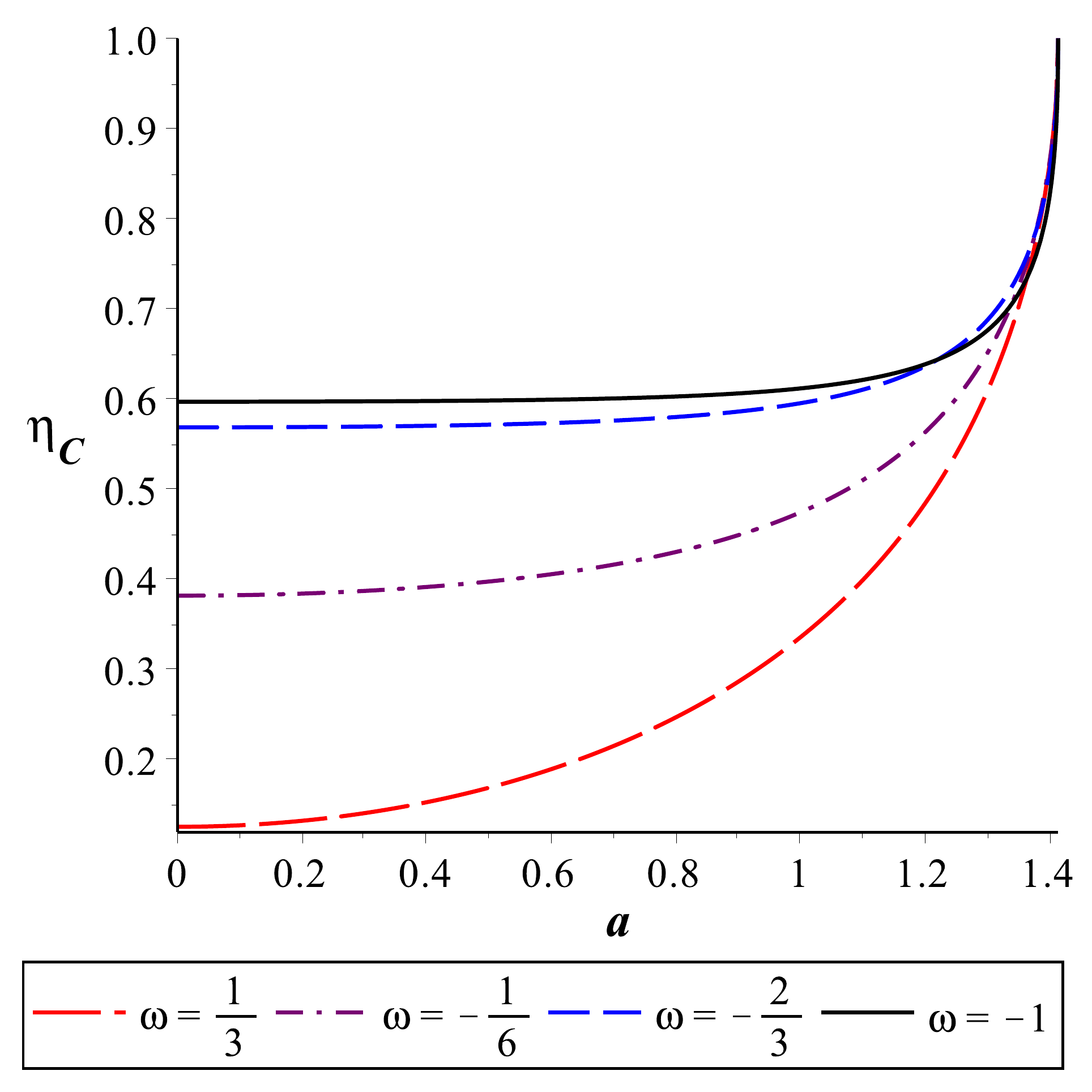}
\caption{Efficiency of the black hole as a Carnot heat engine, $\eta_C$, versus the quantum correction, $a$, for different fluids. Values of the thermodynamic quantities are $\tilde{S}_C=4$, $P_C=0.01$ and $\tilde{S}_H=2$, $P_H=0.05$.}
\label{e-carnot}
\end{figure}

\noindent
where $T_C$ and $T_H$ are the temperatures of the cold and hot reservoirs, respectively, which means that the efficiency cannot be equal to unity, since a reservoir cannot have zero temperature. 

Due to the fact that we have a well-defined formula for the temperature of the reservoir, given by equation (\ref{temperature}), the efficiency of the most efficient black hole can be easily obtained. It ones wants to calculate the efficiency of an arbitrary cycle, then one needs to use equation (\ref{efficiency}).

\begin{figure}[H]
\centering
\includegraphics[scale=0.23]{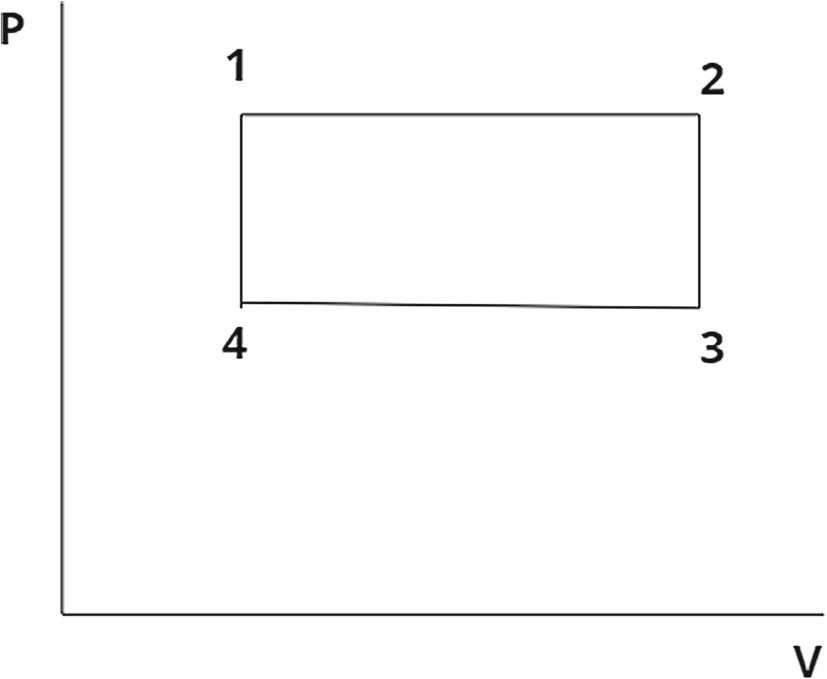}
\caption{A square cycle in the $P-V$ diagram.}\label{square}
\end{figure}

The efficiency for the Carnot cycle is depicted in Fig.(\ref{e-carnot}) for several values of the parameter $\omega$, and as function of the parameter for the quantum correction, $a$. As one can clearly see, the efficiency is improved for all fluids and approaches its maximum as $a$ approaches some fixed value. This fixed value is not absolute but depends on the choices for the entropy and pressure of both reservoirs.  The main concern here is that the introduction of a quantum correction improves the performance of the heat engine.

\begin{figure}[H]
\centering
\includegraphics[scale=0.35]{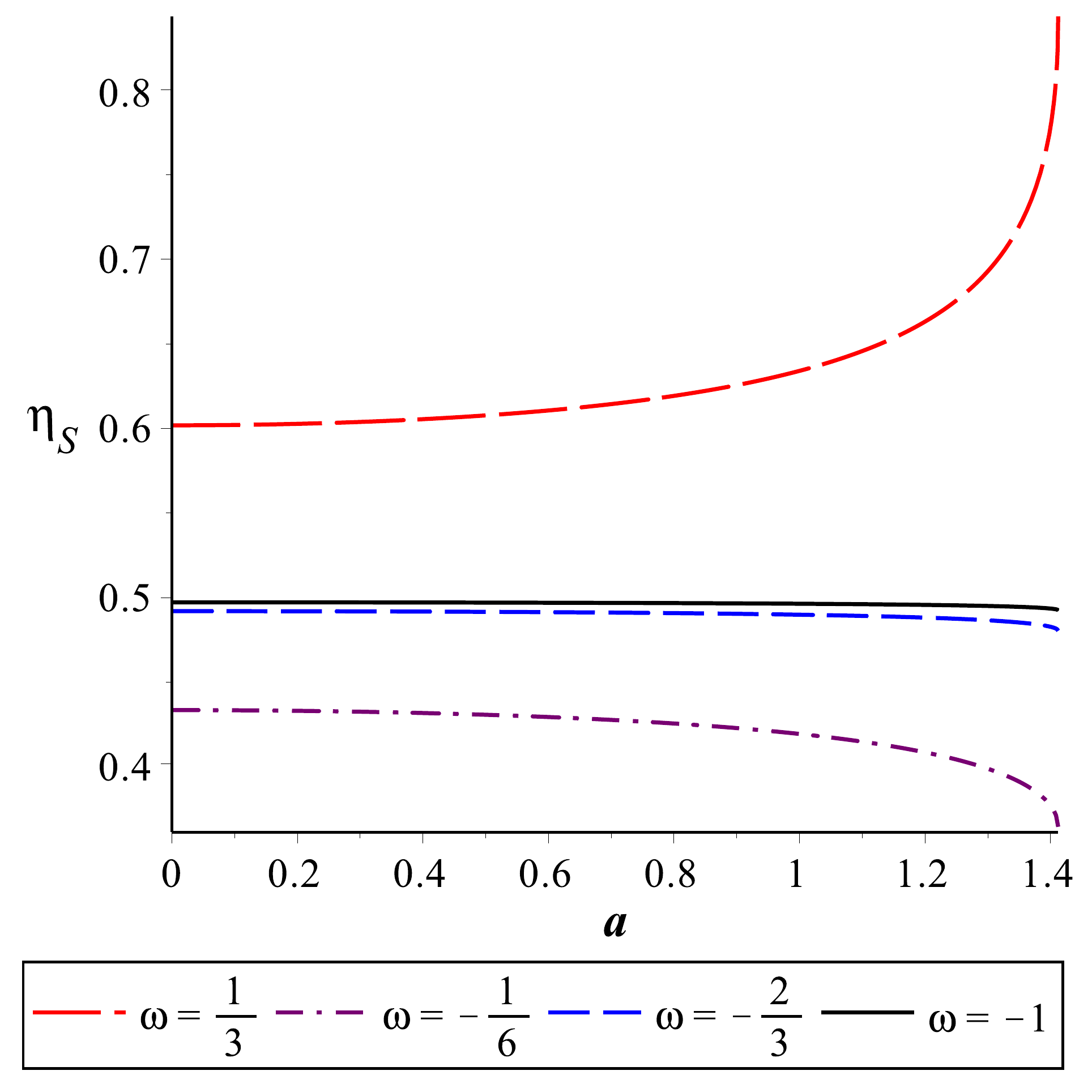}
\caption{Efficiency of the black hole as a heat engine in the square cycle, $\eta_S$, versus the quantum correction, $a$, for different fluids. Values of the thermodynamic quantities are $\tilde{S}_1=\tilde{S}_4=2$, $\tilde{S}_2=\tilde{S}_3=4$ and $P_1=P_2=2$ and $P_3=P_4=1$. We have $M_1>M_2$ and $M4>M_3$.}
\label{e-square}
\end{figure}

Another cycle we can easily study is the square circle, depicted in Fig.(\ref{square}). Since the pressure is constant, the work done by the engine is the variation of the enthalpy, and the enthalpy is the mass of the black hole. This give us a straightforward formula for the efficiency,

\begin{equation}
    \eta = 1 - \frac{M_3 - M_4}{M_2 - M_1},
\end{equation}

\noindent
which is depicted in Fig.(\ref{e-square}). For this cycle, not all kinds of perfect fluids shares the same behaviour as we increase the quantum-correction. For $\omega = 1/3$, the efficiency is improved, but for $\omega = -1/6, -2/3$ and $-1$, the efficiency is reduced, which indicates the different behaviour for normal and exotic matter.

\section{Concluding Remarks}

In this paper, we studied the effect of quantum corrections on the criticality and efficiency of black holes surrounded by a perfect fluid. The main idea was to verify if the system presents first or second order phase transitions due to the introduction of quantum corrections, as compared with the classical black hole.

Although the Schwarzschild metric does not present phase transitions, when one includes quantum corrections this feature is modified and phase transitions can occur. This is a nice feature by itself, since in general, to study phase transitions in a non-rotating black hole, one usually includes electric charge, and so the quantum-corrected black hole is an uncharged black hole where phase transitions can occur. 

We showed that first order phase transitions occur for $\omega>-1/3$ and $w\neq 0$ verified from the existence of saddle points of isotherms in the $p-x$ plane. We also verify a Hawking-Page-like second order phase transition for the cosmological constant case, that is found due to a discontinuity in the derivative of the Gibbs free energy with respect to the temperature, which is a manifestation of a transition that separates low and high mass black holes at a minimum temperature. Second order phase transitions were further explored for other specific fluids and also for a general state parameter.

The study of the black hole as a heat engine also show some interesting features. When we consider the most efficient cycle, the Carnot cycle, the efficiency increases as we increase the parameter of the quantum correction, independently of the equation of state we used. This does not guarantee that the same feature will happens for any equation of state, but at least for radiation, cosmological constant, some other exotic fields, it happens.

When one deals with some arbitrary cycle, the same cannot be stated. For the square cycle, for example, the efficiency of the heat engine does indeed depend on the equation of state, and our study indicates that, for non-exotic fluids, the efficiency increases as we increase the parameter related to the quantum correction, but for the exotic fluids the opposite occurs and the efficiency decreases. 


\acknowledgments
The authors would like to thank CNPq (Conselho Nacional de Desenvolvimento Cient\'ifico e Tecnol\'ogico - Brazil) for financial support. This study was financed in part by the Coordena\c{c}\~ao de Aperfei\c{c}oamento de Pessoal de N\'ivel Superior - Brasil (CAPES) - Finance Code 001.


\end{document}